\documentclass{article}
\usepackage[utf8]{inputenc}
\usepackage{kotex}

\title{RTPS Attack Dataset Description}
\author{School of Cybersecurity in Korea University \\ Hacking and Countermeasure Research Lab \\\\ (Dong Young Kim, Dongsung Kim, Yuchan Song, \\ Gang Min Kim, Min Geun Song, Jeong Do Yoo, Huy Kang Kim)}
\date{November 2023}

\usepackage[numbers]{natbib}
\usepackage{graphicx}
\usepackage{amsmath}
\usepackage{amssymb}
\usepackage{verbatim}
\usepackage{cleveref}
\usepackage{subfig}
\usepackage{tabularx}
\usepackage{titlesec}
\usepackage{makecell}
\usepackage{multirow}
\usepackage{multicol}
\usepackage{booktabs}
\usepackage{adjustbox}
\usepackage{url}
\usepackage{listings}
\usepackage{xcolor}
\usepackage{csquotes}
\definecolor{codegreen}{rgb}{0,0.6,0}
\definecolor{codegray}{rgb}{0.5,0.5,0.5}
\definecolor{codepurple}{rgb}{0.58,0,0.82}
\definecolor{backcolour}{rgb}{0.95,0.95,0.92}

\lstdefinestyle{mystyle}{
    backgroundcolor=\color{backcolour},   
    commentstyle=\color{codegreen},
    keywordstyle=\color{magenta},
    numberstyle=\tiny\color{codegray},
    stringstyle=\color{codepurple},
    basicstyle=\ttfamily\footnotesize,
    breakatwhitespace=false,         
    breaklines=true,                 
    captionpos=b,                    
    keepspaces=true,                 
    numbers=left,                    
    numbersep=5pt,                  
    showspaces=false,                
    showstringspaces=false,
    showtabs=false,                  
    tabsize=2
}
\titleformat{\paragraph}
{\normalfont\normalsize\bfseries}{\theparagraph}{1em}{}
\titlespacing*{\paragraph}
{0pt}{3.25ex plus 1ex minus .2ex}{1.5ex plus .2ex}

\begin{document}

\maketitle

% 김동영
\begin{abstract}
We collect Attack and Normal packet by injecting attack data in
an Unmanned Ground Vehicle (UGV) in a normal state. We assembled a testbed for the UGV, controller, PC, and router to collect this dataset. We conducted two types of attacks on the testbed: Command Injection and ARP Spoofing. The data collection times are 180, 300, 600, and 1,200. The scenario has 30 each at the collection time, 240 total. We hope this dataset will contribute to developing technologies such as anomaly detection to address security threat issues in ROS2 networks and UGVs.
\end{abstract}


\section*{Acknowledgment}
\noindent This work was supported by the Institute for Information \& Communications Technology Promotion (IITP) grant funded by the Korea government (MSIT). (Grant No. 2020-0-00374, Development of Security Primitives for Unmanned Vehicles).

\begin{figure}[b]
    \centering
    \includegraphics[height=0.6in]{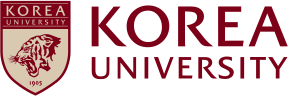}
    \hspace{1cm}
    \includegraphics[height=0.6in]{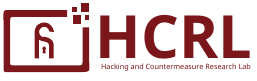}
\end{figure}
\onecolumn\tableofcontents

\newpage
\section{Abbreviations and acronyms} \label{sec:abbreviations}
This document uses these abbreviations and acronyms:
% 알파벳 순 정렬
\begin{description}
    \item[AHRS] Attitude Heading Reference System
    \item[DDS] Data Distribution Service
    \item[LiDAR] Light Detection And Ranging
    \item[PDB] Power Distribution Board
    \item[ROS] Robot Operating System
    \item[RTPS] Real Time Publish Subscribe Protocol
    \item[SBC] Single Board Computer
    \item[UGV] Unmanned Ground Vehicle
    \item[VM] Virtual Machine
\end{description}

% 송민근
\newpage
\section{Introduction} \label{sec:intro}

ROS2 is an open-source meta-operating system for robot development and control, providing developers and researchers with essential tools and libraries. As the next-generation version of ROS, ROS2 has been developed to improve security and reliability, making robot operations in real-world environments more stable.

ROS2 is used in various applications such as autonomous vehicles, industrial equipment, and robotics, increasing its significance in security. Consequently, some research is conducted to understand the types of potential attacks, identify security vulnerabilities, and develop appropriate countermeasures in ROS2 \cite{kim2018security, deng2022security}.

\begin{sloppypar}
In this paper, we introduce the ROS2-based RTPS attack dataset. This dataset consists of data collected from various attack scenarios incorporating various data collection times, attack durations, and rest periods. It will help to analyze the attack patterns in RTPS or related packet. We expect this dataset will support researchers to investigate the vulnerabilities in ROS2 system and utilize it as fundamental material for future security improvements.
\end{sloppypar}

This document aims for professionals and academic researchers conducting studies on network security in autonomous robots. This paper is intended to allow readers to understand the structure of the dataset, attack scenarios, and labeling methods.

This paper is divided into seven sections. Section~\ref{sec:abbreviations} details the abbreviations used in this document. Section~\ref{sec:intro} is the Introduction to the paper and provides an overview of the dataset. Section~\ref{sec:background} comprehensively explains ROS1 and ROS2 and the RTPS Protocol used in ROS2. Section~\ref{sec:testbed} describes the testbed environment used for data extraction in the dataset. Section~\ref{sec:methodology} explains the attack methods used in the dataset. Section~\ref{sec:attackscenario} discusses attack scenarios. Lastly, Section~\ref{sec:metadata} describes the information in the file about the dataset.

% 송민근
\newpage
\section{Background} \label{sec:background}
\subsection{ROS (Robot Operating System)}
\begin{sloppypar}
ROS1 and ROS2 are open source Robot Operating Systems for developing, deploying, and operating robots. While these two systems share similar objectives, ROS2 has been developed as the next-generation version of ROS1 and features several key differences \cite{ROS1, ROS2}.
\end{sloppypar}

\begin{itemize}
    \item ROS1:
    \begin{itemize}
        \item Open Source: It is developed as community-based open-source software and is widely used among robot developers and researchers.
        \item Distributed architecture: It supports modularization and distributed communication in robot systems, allowing the development and execution of robot software across multiple nodes.
        \item C++ and Python Support: It supports various programming languages, including C++ and Python, facilitating robot software development.
        \item Mature Ecosystem: It provides a large amount of libraries, plugins, and modules that support integration with various robot hardware and sensors.
        \item Community support: It is supported by an active community of developers and users, offering a variety of documentation and tutorials.
        \item Real-time Operation Support: It is designed to function in real-time applications, allowing real-time execution of robot control and sensing applications.
    \end{itemize}

    \item ROS2:
    \begin{itemize}
        \item Enhanced Security and Reliability: It offers more robust security and reliability than ROS1. Moreover, it is developed to stabilize robot operations in real environments.
        \item Improved Distributed Architecture: It performs better in distributed systems, facilitating communication and collaboration between multiple robot nodes.
        \item Multi-Language Support: It supports various languages, including Python, C++, and Java, providing more options for robot software development.
        \item Real-time Operation Support: It is designed to operate in real-time applications.
        \item Standardized Communication Protocol: It uses the DDS \cite{dds_specification} protocol to support reliable data communication.
    \end{itemize}
\end{itemize}

\begin{sloppypar}
ROS2 has been improved over ROS1 in numerous aspects, such as security, performance, and support for distributed systems. Therefore, a large number of robot developers and researchers prefer ROS2.
\end{sloppypar}

\subsection{RTPS Protocol}
RTPS \cite{specification2007real} protocol is commonly used in industrial devices for its communications. With the increasing use of Data DDS in industrial systems, RTPS was developed as a standard protocol to address compatibility issues and others. It is a protocol for time-sensitive data, mainly used in real-time and embedded systems. RTPS enables reliable data exchange between multiple systems through Publish and Subscribe connections among nodes. It is commonly utilized with DDS to implement the data transmission and communication mechanisms of DDS.

The following are the key features of the RTPS protocol:
\begin{itemize}
    \item Publish-Subscribe Model: It utilizes the Publish-Subscribe model for communication between data publishers and subscribers. Data Publishers transmit data, while Data Subscribers receive it.
    \item Distributed architecture: It supports a distributed architecture for data exchange between multiple systems. It operates reliably even when various systems are connected via a network.
    \item Transparent Data Transmission: Allows for transparent data transmission across networks. In other words, transmission is reliable even when Data Publishers and Subscribers are on different systems.
    \item Quality of Service (QoS) Support: It provides QoS features to regulate the quality of data transmission. It fulfills requirements like reliability, bandwidth, and priority in real-time systems.
    \item Dynamic Registration of Publishers and Subscribers: It supports the dynamic registration and deregistration of data Publishers and Subscribers. It is utilized when data publishers or Subscribers are added or removed from the system.
    \item Detection and Discovery: It offers the ability to detect and discover other data Publishers and Subscribers on the network. It facilitates smooth data exchange even in dynamically changing network environments.
    \item Support for Real-Time Operation: It is designed for real-time applications that transmit and receive data in real-time.
\end{itemize}

RTPS is used in distributed system platforms, and when used in conjunction with DDS, it provides high levels of data reliability and real-time performance. It is primarily utilized in various real-time applications such as embedded systems, robotics, automotive, telecommunications equipment, and industrial control systems.

% 김동영
\section{Testbed} \label{sec:testbed}

\subsection{Testbed Environment}

\begin{figure}[ht!]
    \includegraphics[width=\linewidth]{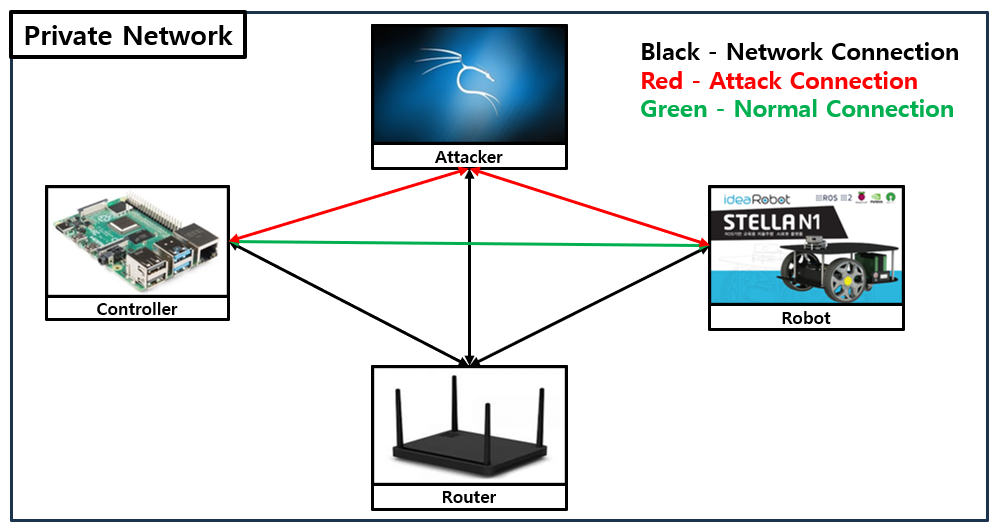}
    \caption{Structure of the testbed for dataset Collection}
    \label{fig:ros_network}
\end{figure}

In this section, We will describe the testbed. It is used for extracting the dataset. It consists of an Attacker, a Controller, a Robot (Stella N1), and a Router. Figure~\ref{fig:ros_network}. shows the environment's structure for collecting the dataset. The Controller and the Robot communicated within a single Router using communication protocols. In this scenario, the attacker spoofed into the internal network to launch attacks, and the attack dataset was collected.  

We thought about not mixing with other packet made out of the testbed. The router wouldn't connect to the Internet (No external connection). It prevents the collection of collection packets generated by devices not part of the testbed.

\newpage
\subsection{Hardware}
The components of the testbed used in this study are as follows:

\begin{itemize}
    \item Victim Robot (Stella N1)
    \item Controller (Raspberry Pi 4)
    \item Attacker PC (Kali Linux VM)
    \item Router
\end{itemize} 

\subsubsection{Victim Robot} \label{sec:victim_robot}

\begin{figure}[ht!]
\includegraphics[width=\linewidth]{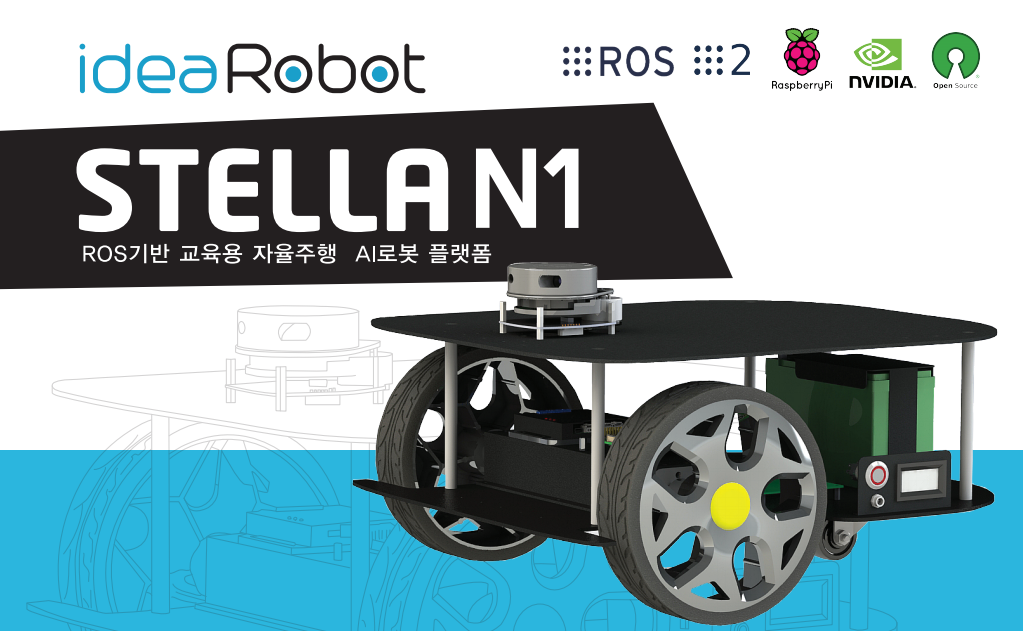}
    \caption{Stella N1 \cite{stella_n1_image}}
    \label{fig:stella_n1}
\end{figure}

In this experiment, the robot that was victim subjected to attacks was Stella N1~\cite{stella_n1}, a ROS2-using robot, as shown in Figure~\ref{fig:stella_n1}. This product was equipped with ROS2, and Fast-DDS uses RTPS as its communication protocol. In this document, we explain a scenario constructed where attacks are carried out on the Stella N1 via the RTPS protocol by an attacker to collect a dataset.

\begin{figure}[ht!]
\includegraphics[width=\linewidth]{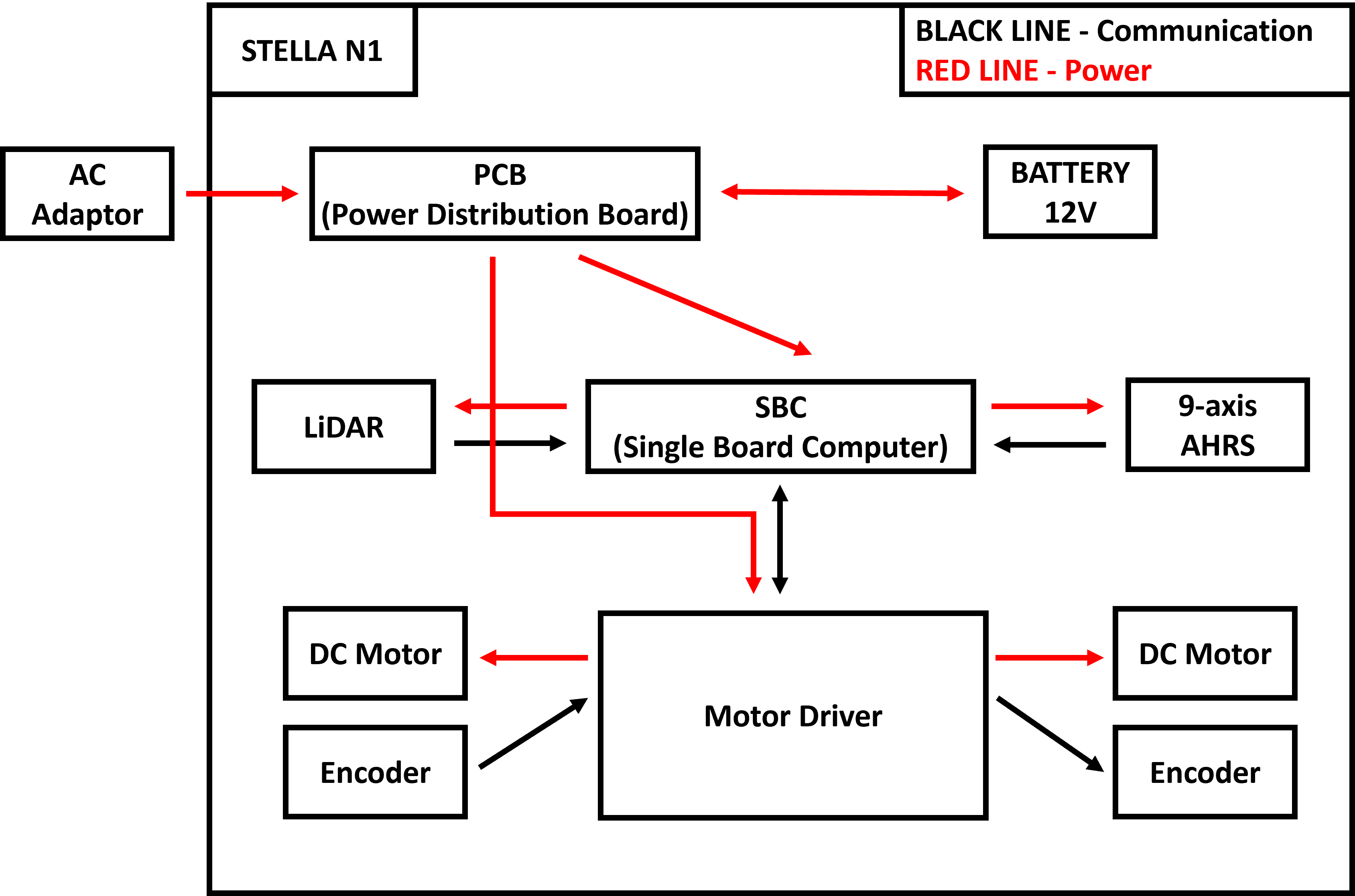}
    \caption{Stella N1 block diagram \cite{stella_block_diagram}}
    \label{fig:stella_n1_diagram}
\end{figure}

\begin{sloppypar}
Figure~\ref{fig:stella_n1_diagram} shows the block diagram of the robot used in the experiment. The robot was equipped with a PDB for power distribution, an AHRS for coordinate system functions, a LiDAR for scanning using autonomous driving, a DC motor for robot propulsion, and a Raspberry Pi for communication with the controller. \\
\end{sloppypar}

The specifications and significant components of Stella N1 are as follows~\cite{stella_n1_parts}:

\begin{itemize}
    \item SBC: Operates ROS2 to control the robot.
    \begin{itemize}
        \item Product Name: Raspberry Pi 4 Model B
        \item Processor: Broadcom BCM2711, quad-core Cortex-A72 (ARM v8) 64-bit SoC @ 1.5GHz 
        \item Memory: 4GB with on-die ECC
        \item Connectivity: 2.4 GHz and 5.0 GHz IEEE 802.11b/g/n/ac wireless
        \item Operating System (OS) Configuration:
        \begin{itemize}
            \item Host OS: Raspbian (x64)
            \item ROS Version: ROS2 Foxy
        \end{itemize}
    \end{itemize}
    \item PDB: Responsible for power distribution
    \item AHRS: Serves as the coordinate system for the robot
    \item LiDAR: Performs scanning of the robot's surroundings
    \item DC Motor: Drives the robot's wheels
    \item Encoder: Detects mechanical positional changes
    \item Motor Driver: Controls the DC Motor
\end{itemize}

\subsubsection{Controller}

The controller was based on Ubuntu Server 20.04 with Desktop Version and uses ROS2 Foxy. The controller had the role of sending commands to operate the robot and was considered a legitimate user. In the case of a Command Injection attack, the attack code was executed within the controller.
\\
The specific specifications of the controller are as follows:
\begin{itemize}
    \item Product Name: Raspberry Pi 4 Model B
    \item Processor: Broadcom BCM2711, quad-core Cortex-A72 (ARM v8) 64-bit SoC @ 1.5GHz 
    \item Memory: 4GB with on-die ECC
    \item Connectivity: 2.4 GHz and 5.0 GHz IEEE 802.11b/g/n/ac wireless
    \item OS Configuration:
        \begin{itemize}
            \item Host OS: Ubuntu Server 20.04 with Desktop Version (x64)
            \item ROS Version: ROS2 Foxy
        \end{itemize}
\end{itemize}

\subsubsection{Attacker PC}

\begin{sloppypar}
The attacker's PC was based on Windows 11. The attack came from a Kali Linux VM. The attacker's PC was only used in the ARP Spoofing attack scenario and not used in the Command Injection scenario.
\end{sloppypar}

The specific specifications of the attacker's PC are as follows:
\begin{itemize}
    \item Processor: Intel(R) Core(TM) i7-9700K CPU @ 3.60GHz
    \item Memory: 32GB with DDR4
    \item Connectivity: 2.4 GHz and 5.0 GHz IEEE 802.11b/g/n/ac wireless
    \item OS Configuration:
        \begin{itemize}
            \item Host OS: Windows 11 Pro 22H2 22621.2215
            \item Virtual Environment: VMware WORKSTATION Pro 17.0.0 build-20800274
            \item Virtual OS: Kali Linux 2023.3 (x64)
        \end{itemize}
\end{itemize}

% 김동영
\newpage
\section{Methodology} \label{sec:methodology}

\subsection{Attack Types}
\begin{figure}[ht!]
\includegraphics[width=\linewidth]{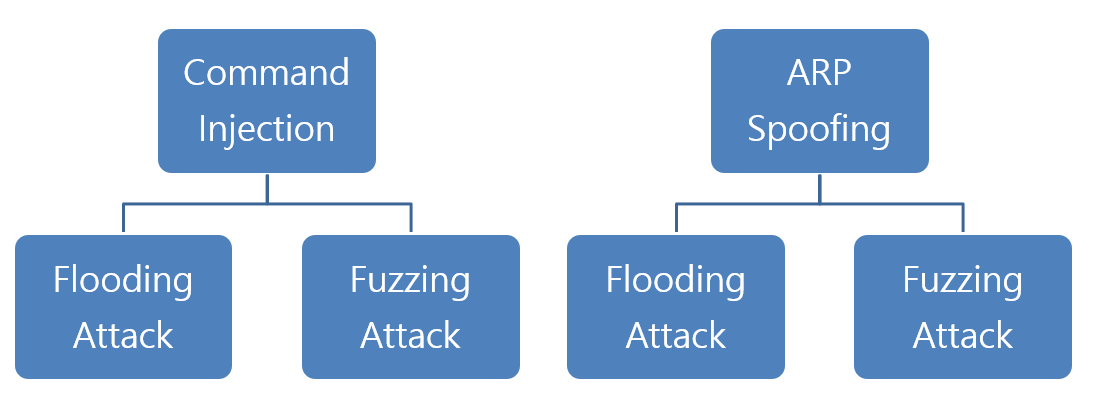}
    \caption{Two types of attacks on dataset}
    \label{fig:attack_method}
\end{figure}

In the experiment to extract this dataset, there were two injection attack methods on the RTPS protocol used in autonomous vehicles, as shown in Figure~\ref{fig:attack_method}. These attacks involved sniffing the network from a controller operating generally to the vehicle and then extracting only the key field values from the packet controlling the robot. Subsequently, new RTPS packets were created using the manipulated values of these extracted fields and the acceleration value, SerializedData. These methods are differentiated as Command Injection where SerializedData is altered and a combined method where ARP Spoofing is also performed on the Command Injection attack. Each attack included a Flooding Attack, which sent fixed data continuously, or a Fuzzing Attack, which continuously sent random numbers.

\subsection{Command Injection} \label{sec: command_injection}
The attacker sniffed packets by controlling the robot, copying values such as checksums. Afterward, the copied RTPS packet was generated, and the desired SerializedData value (attacker's defined value) was inserted. The structure of SerializedData varied for each application, and this dataset used the robot described in Section~\ref{sec:victim_robot}. The data to be included in SerializedData was written by referencing the robot's Publisher code~\cite{stella_teleop_key}.

In the case of a Flooding Attack, involved injecting a large amount of data that performs specific actions. For collecting these data, the attacker performed an attack by generating a large amount of data that exists in a stopped state, that is, data with a speed of 0, to cause abnormal behavior. Therefore, a dataset containing normal and abnormal data was created.

In the case of a Fuzzing Attack, random data was injected to induce abnormal behavior. The goal was to make the robot exhibit unpredictable behavior, similar to normal and abnormal datasets.

\subsection{ARP Spoofing}

\begin{figure}[ht!]
\includegraphics[width=\linewidth]{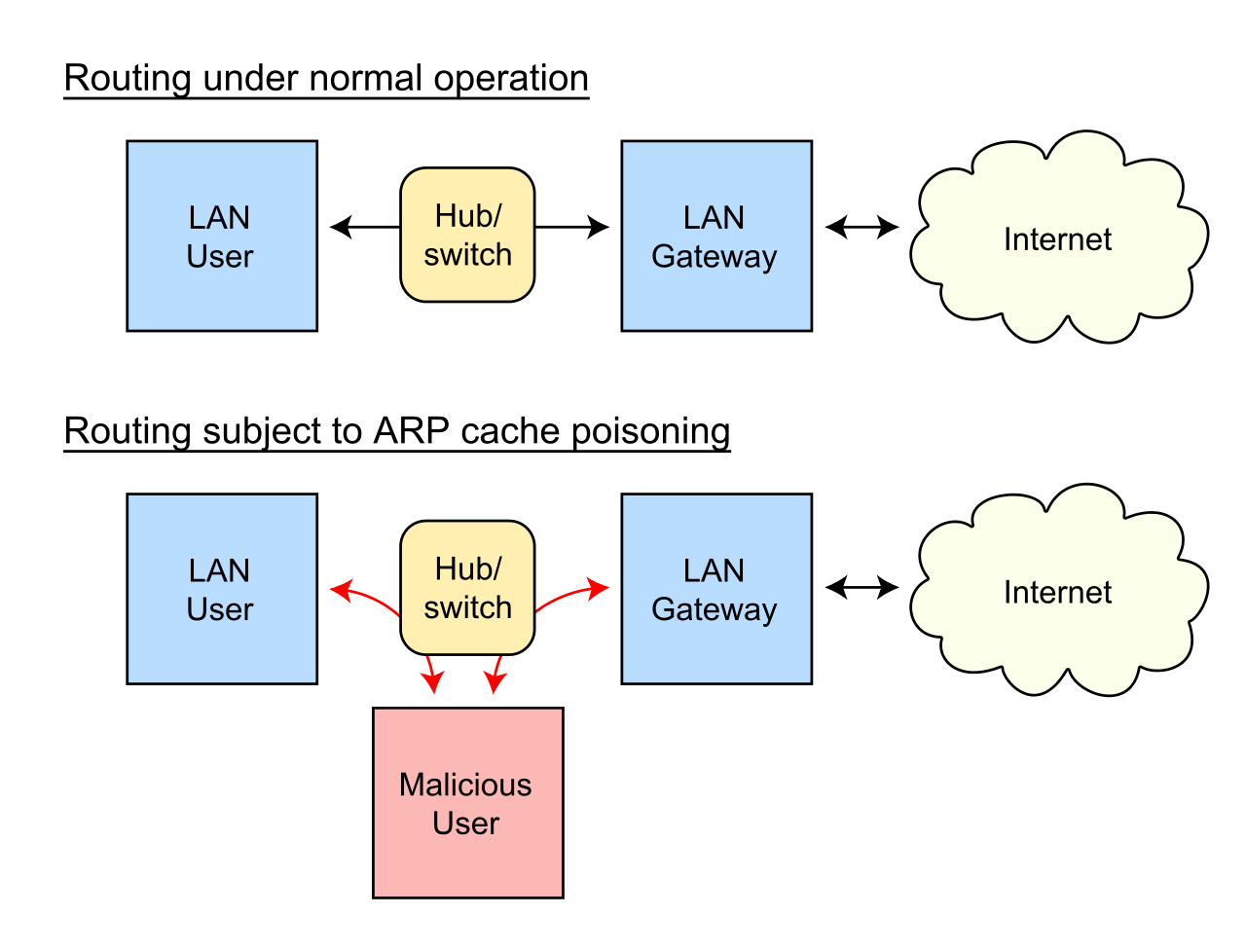}
\caption{Difference between regular and ARP Spoofing Communication \cite{arp_spoofing}}
\label{fig:arp_spoofing}
\end{figure}

ARP Spoofing was an attack technique in which an attacker intercepted data packets in communication between normal users and robots. By manipulating the robot's MAC table, the attacker replaced the controller's MAC address with the attacker's MAC address. The attacker tricked the normal user into sending data, believing that the attacker was the legitimate controller, and manipulated the data packets sent by the attacker to perform desired actions. The difference between ARP Spoofing and regular communication is depicted in Figure~\ref{fig:arp_spoofing}.

This experiment used ARP Spoofing to make the robot recognize the attacker as a normal user and perform flood and fuzzing attacks as described in Section~\ref{sec: command_injection}. ARP Spoofing started from the beginning of data collection and ended when data collection was complete. Therefore, the data from the ARP Spoofing, Flooding, and Fuzzing attack packets were mixed with the normal data in the dataset. In this experiment, we used arpspoof from dnsiff~\cite{arp_spoof_tool}, executed on a Kali Linux VM to carry out the attack.

% 김강민
\section{Attack Scenarios} \label{sec:attackscenario}

In this section, we discuss the anticipated effects of attacking autonomous vehicles and the methods used to collect the dataset.

\subsection{Attack Expectations}

The dataset was created specifically for autonomous vehicles. The attacks performed were Command Injection and ARP Spoofing. Command Injection can be further divided into Spoofing attacks, where arbitrary commands are injected, and Fuzzing attack, here random commands are injected. These attack can compromise the command system of autonomous vehicles, as discussed in Section~\ref{sec:methodology}.

The dataset comprises packets targeting the RTPS system, utilized in the open-source middleware ROS2 and DDS. Therefore, it could enhance the security of programs utilizing RTPS in ROS and DDS. While the dataset was created specifically for autonomous vehicles, it communicates through the SerializedData of Data Submessages and can contribute to network security for all programs using unencrypted RTPS packets.

\subsection{Attack Scenarios Composition}

\begin{figure}[!htb]
    \centering
    \subfloat[Scenarios sequence: attack\_by\_once]{
    \includegraphics[width=0.95\linewidth]{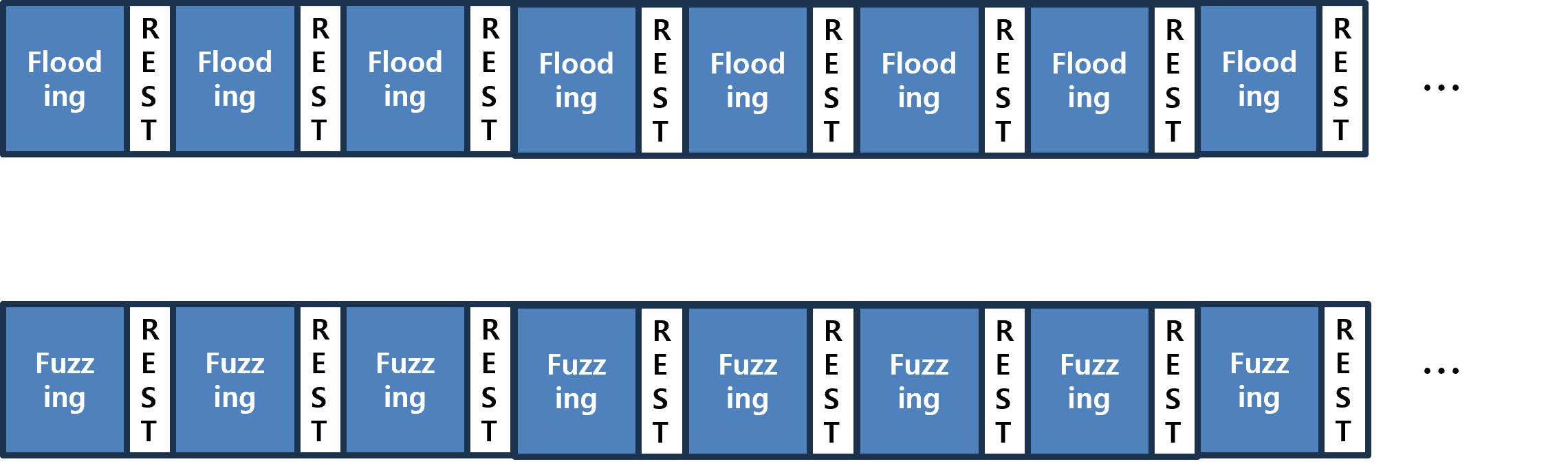}
        \label{fig:attack_by_once}
    }\\ % Changed \linebreak to \\ for better compatibility
    \subfloat[Scenarios sequence: attack\_by\_shuffled]{
    \includegraphics[width=0.95\linewidth]{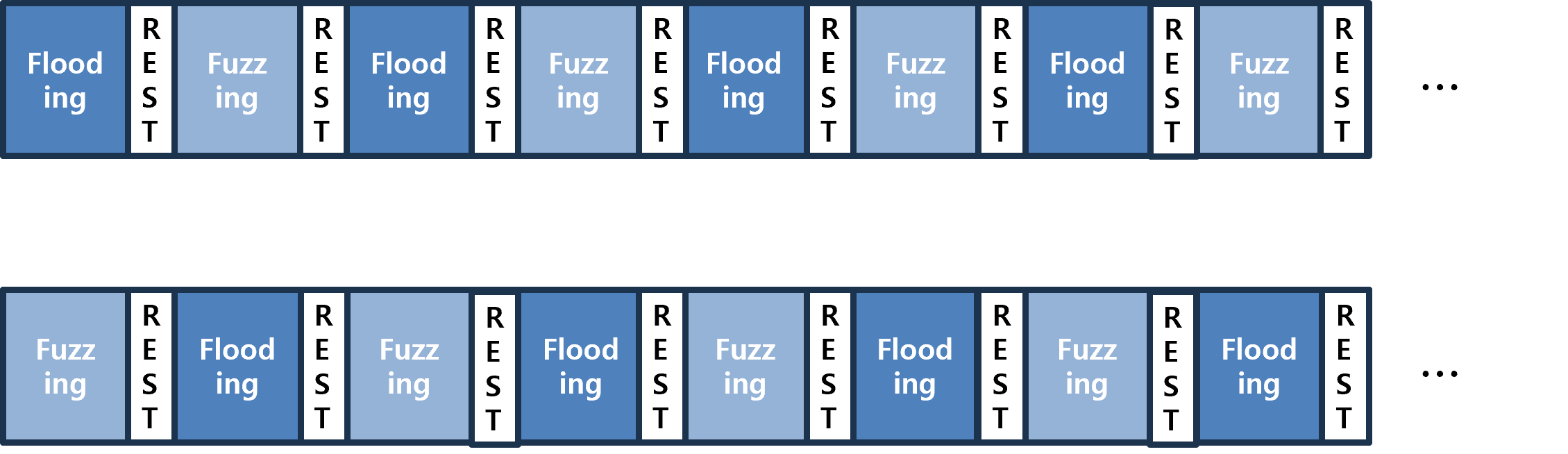}
        \label{fig:attack_by_shuffled}
    }\\ % Changed \linebreak to \\ for better compatibility
    \subfloat[Scenarios sequence: attack\_by\_random]{
    \includegraphics[width=0.95\linewidth]{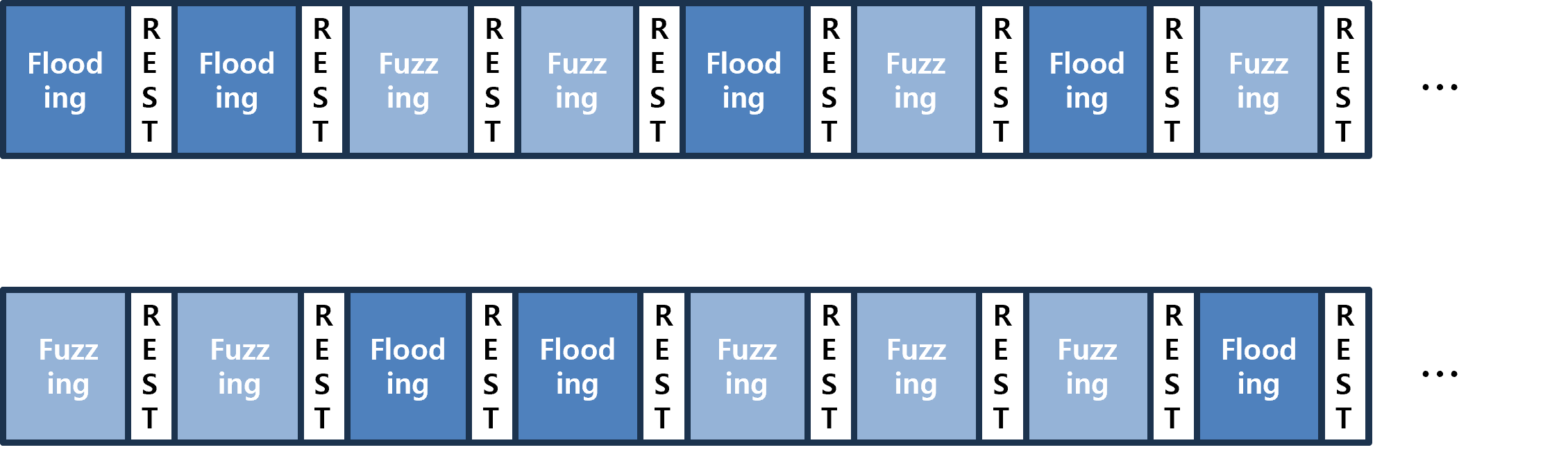}
        \label{fig:attack_by_random}
    }\\ % Added for consistency
    \caption{Overview of Scenario Sequence}
    \label{fig:attack_order}
\end{figure}

The dataset comprises various scenarios combining data collection time, attack sequence, and attack method. Data collection time refer to the unit in which data are collected and organized into folders with time intervals of \textit{180, 300, 600, 1,200} seconds. There are two type of Attack method \textit{Command Injection attacks} and \textit{ARP Spoofing + Command Injection attacks}. Each attack includes subattack method, which is \textit{Flooding} and \textit{Fuzzing}. Attack sequence denotes the order in which the attack methods are mixed, as shown in Figure~\ref{fig:attack_order}:

\begin{itemize}
    \item Figure \ref{fig:attack_by_once}. One attack is carried out (attack\_by\_once)
    \item Figure \ref{fig:attack_by_shuffled}. Proceeding by mixing multiple attacks (attack\_by\_shuffled)
    \item Figure \ref{fig:attack_by_random}. Proceeding by randomly determining the attack method (attack\_by\_random)
\end{itemize} 

The attack sequence is composed by combining Rest Time and Attack Time. Rest time indicates the duration of rest after each attack, with time intervals of \textit{10 and 20} seconds. Attack Time represents the duration in which the Flooding or Fuzzing attacks are executed, with time intervals of \textit{10, 50, and 100} seconds. 

\begin{figure}[ht!]
\includegraphics[width=\linewidth]{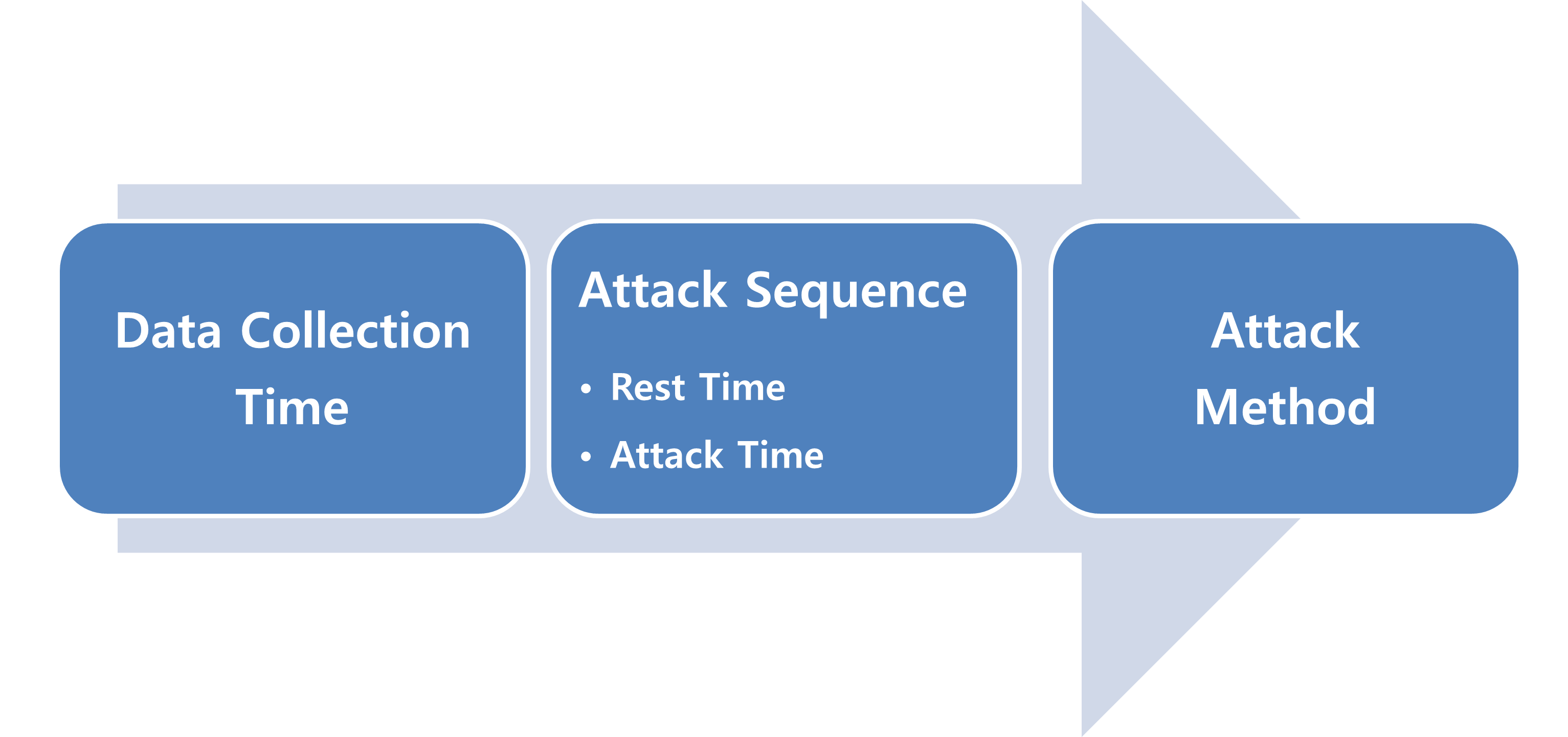}
    \caption{Whole procedure for collecting dataset}
    \label{fig:dataset_scenario_structure}
\end{figure}

Based on the aforementioned information, the data collection tool is structured using nested loops corresponding to Data collection time, Attack sequence (Rest time and Attack time), and Attack method. Figure~\ref{fig:dataset_scenario_structure} visualizes this structure. By utilizing this code, a total of 30 scenarios and datasets were generated for each data collection time.

\subsection{Collecting the Dataset}

The dataset was acquired via automated code execution, encompassing attacks, equipment manipulation, and data collection. It consists of code for automating data collection and attacks, and for simulating robot control. The execution flow for each code is as follows:

\begin{itemize}
    \item main.py
\\ The dataset automation code starts by executing of this script. In the case of ARP Spoofing attacks, it calls a function to automate ARP spoofing attacks. It then reboots the robot and determines the data collection time. If the remaining execution time exceeds the attack time, it forcefully rests. It also calls the automation function for each attack scenario based on its type.
    
    \item Automating data collection and attack code
\\ main.py calls the scenario automation function for each attack type. When a scenario automation function starts, it measures the start time and calls a function to remotely command the robot to collect packets. It then calls the attack function based on the predetermined attack and rest times, as well as the attack method, initiating the attack or taking rest accordingly. Once a scenario is completed, it transfers the captured packet dump files in the robot to the controller or the attacker's PC. Finally, it concludes the data collection code for that scenario.
    
    \item Simulating the Operation of a Robot Code
\\ The robot used to collect this dataset provides control functionality using a keyboard. Commands such as \textit{w} for forward, \textit{a} for left turn, \textit{d} for right turn, \textit{x} for reverse, and s for complete stop are provided. Each key press is implemented as a function, with a random number generation function called to input \textit{a} key. For example, if a random number between \textit{1} and \textit{14} is generated, an input of \textit{1} corresponds to moving forward.
\end{itemize}

The above code obtains packet dump files (.pcap) from the robot and packet dump files collected at the time of the attack (labeled.pcap). They are organized sequentially from 0 to 29 for each scenario.

\newpage
\subsection{Labeling the Dataset} \label{sec:labeling}

We labeled the dataset for Command Injection and ARP Spoofing.

For Command Injection, we verified if the packet from the robot matched the packet during the attack and checked if the packet was an RTPS packet. Then We proceeded to analyze by comparing the Timestamp value of the \textit{INFO\_TS Submessage}, the \textit{writerEntity} value of the \textit{Data Submessage}, the UDP checksum, and the packet’s source and destination IP addresses. If the collected packet matched the attack packet, we labeled it as Attack.

For ARP Spoofing, we first checked if the packet was an ARP packet and then verified if the \textit{ARP Opcode} value was 2, which indicates an ARP Reply packet that changes the MAC address. We also evaluated if the Source MAC address was the same as the attacker's MAC address. If these conditions were met, we labeled the packet as Attack.

\begin{figure}[ht!]
\includegraphics[width=\linewidth]{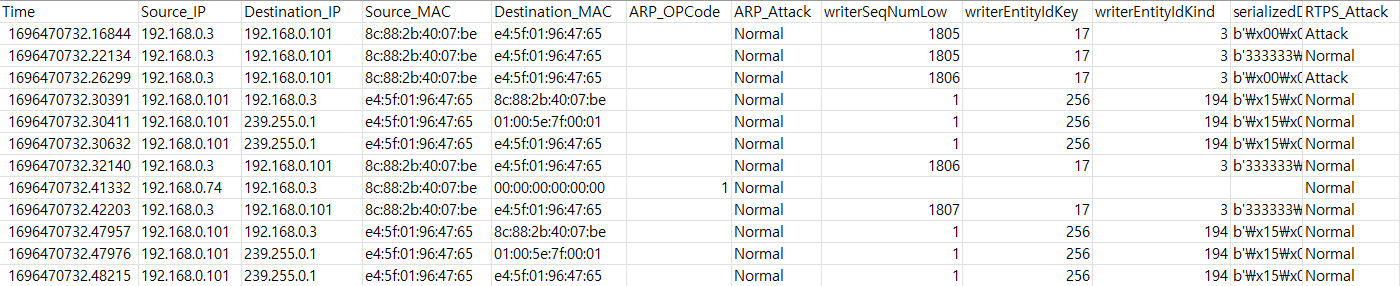}
    \caption{Part of labeling example file}
    \label{fig:label_example}
\end{figure}

Figure~\ref{fig:label_example} shows a portion of a labeled example file. If a particular value didn't exist, we set it as \textit{None} to appear as a space.

\begin{table}
\centering
\begin{tabular}{|c|c||c|c|} 
\hline
Number of files & Quantity & Size of files & Size (MB)
\\ \hline \hline
Robot Packet Dump & 240 & Robot Packet Dump & 1,068.2 \\ \hline
Attack Packet Dump & 2,948 & Attack Packet Dump & 124.9 \\ \hline
Labeling Files & 8 & Labeling Files & 3,148.7 \\ \hline
Total Files & 3,196 & Total Size & 4,341.8 \\ \hline
\end{tabular}
\caption{Metadata of the dataset}
\label{tab:meta_data} 
\end{table}

\section{Dataset Information} \label{sec:metadata}

In this section, we describe the main information of the dataset, which collected normal data and attack data that occurred after the scenario attack of Section~\ref{sec:attackscenario}.

\subsection{Information of the Dataset}

Table~\ref{tab:meta_data} provides information about the dataset files. There are 240 robot packet dump files with a total size of 1,008.2 MB. Attack packet dump
files consist of 240 files with a total size of 124.9 MB. The dataset includes 8 labeled files with a total size of 3,148.7 MB. Finally, the total number of files in the dataset is 3,196, with a total size of 4,341.8 MB.

The robot packet dump files are in the .pcap file format and store all packet during the scenario-specific data collection process. The attack packet dump files are also in the .pcap file format and store the packet captured during the attack. The labeled files are in .csv file format and contain data similar to what is shown in Figure~\ref{fig:label_example}. The labeling process was explained in detail in Section~\ref{sec:labeling}.

\subsection{Structure of dataset}

\begin{figure}[ht!]
\includegraphics[width=\linewidth]{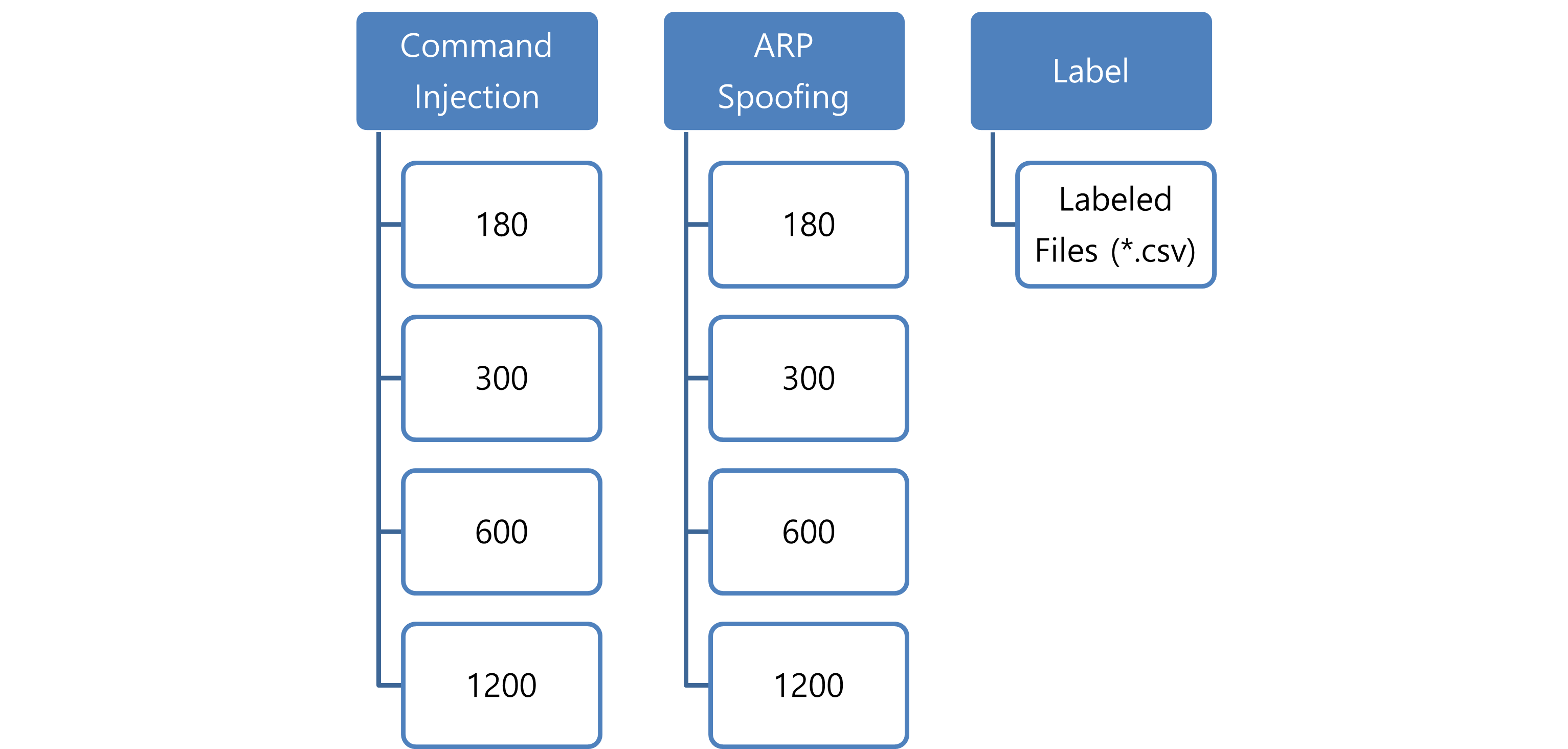}
    \caption{Structure of dataset folders}
    \label{fig:dataset_file_structure}
\end{figure}

The dataset folders are organized as shown in Figure~\ref{fig:dataset_file_structure}. The dataset comprises robot attack files (*.pcap), files representing the packets recorded when the attack occurred (*labeled.pcap), and corresponding labeled files (*.csv). These files contain packet received by the robot during an attack, packet captured when the attacker performs the attack, and labeled information for the respective data.

The packet files included in the dataset are organized into folders based on the collection time, with each folder containing data collected for 180 , 300, 600, or 1,200 seconds. These files include packet captured for the duration of the attack.

\begin{sloppypar}
The .csv files contain labeled information about ARP Spoofing and Command Injection. The labeling is performed using the methodology described in Section~\ref{sec:labeling} to accurately identify the attack types. The files include various pieces of information, with key elements such as Time, Source IP, and Destination IP. Additionally, it includes information such as Source MAC, Destination MAC, and ARP OPCode, which can be useful for developing attack detection systems.
\end{sloppypar}

\bibliography{main}

\begin{thebibliography}{10}

\bibitem{kim2018security}
Jongkil Kim, Jonathon~M Smereka, Calvin Cheung, Surya Nepal, and Marthie Grobler.
\newblock Security and performance considerations in ros 2: A balancing act.
\newblock {\em arXiv preprint arXiv:1809.09566}, 2018.

\bibitem{deng2022security}
Gelei Deng, Guowen Xu, Yuan Zhou, Tianwei Zhang, and Yang Liu.
\newblock On the (in) security of secure ros2.
\newblock In {\em Proceedings of the 2022 ACM SIGSAC Conference on Computer and Communications Security}, pages 739--753, 2022.

\bibitem{ROS1}
Open Robotics.
\newblock Ros introduction.
\newblock \url{https://wiki.ros.org/ROS/Introduction}, 2022.
\newblock Accessed: 2023-11-24.

\bibitem{ROS2}
Open Robotics.
\newblock Ros 2 documentation: Rolling.
\newblock \url{https://docs.ros.org/en/rolling/The-ROS2-Project/Feature-Ideas.html}, 2023.
\newblock Accessed: 2023-11-24.

\bibitem{dds_specification}
OMG Specification.
\newblock Omg data distribution service (dds).
\newblock {\em Object Management Group Pct15-04-10}, 2015.

\bibitem{specification2007real}
OMG Specification.
\newblock The real-time publish-subscribe protocol (rtps) dds interoperability wire protocol specification.
\newblock {\em Object Management Group Pct07-08-04}, 2007.

\bibitem{stella_n1_image}
idea robot.
\newblock Stella n1 user manual.
\newblock \url{https://3187770235-files.gitbook.io/~/files/v0/b/gitbook-legacy-files/o/assets%2F-MhvrKjF6_60tNjTx2_K%2F-MhvrZhPip2b_OQitZ6v%2F-MhvvY9037h8eCBDsiqo%2FSTELLA%20N1.png?alt=media&token=43627b93-af83-43d7-91a5-61ec79b1e839}, 2023.
\newblock Accessed: 2023-11-24.

\bibitem{stella_n1}
idea robot.
\newblock Stella n1 user manual.
\newblock \url{https://idearobot.gitbook.io/stella-n1}, 2023.
\newblock Accessed: 2023-11-24.

\bibitem{stella_block_diagram}
idea robot.
\newblock Stella n1 block diagram.
\newblock \url{https://idearobot.gitbook.io/stella-n1/stella-n1/undefined-2}, 2023.
\newblock Accessed: 2023-11-24.

\bibitem{stella_n1_parts}
idea robot.
\newblock Stella n1 parts specification.
\newblock \url{https://idearobot.gitbook.io/stella-n1/stella-n1/undefined-3}, 2023.
\newblock Accessed: 2023-11-24.

\bibitem{stella_teleop_key}
ntrexlab.
\newblock stella\_teleop\_key.py.
\newblock \url{https://github.com/ntrexlab/STELLA_REMOTE_PC/blob/main/stella_teleop/src/stella_teleop_key.py}, 2021.
\newblock Accessed: 2023-11-24.

\bibitem{arp_spoofing}
Wikipedia.
\newblock File:arp spoofing.svg.
\newblock \url{https://ko.wikipedia.org/wiki/ARP_%EC%8A%A4%ED%91%B8%ED%95%91#/media/%ED%8C%8C%EC%9D%BC:ARP_Spoofing.svg}, 2011.
\newblock Accessed: 2023-11-24.

\bibitem{arp_spoof_tool}
Dug Song.
\newblock dsniff.
\newblock \url{https://www.monkey.org/~dugsong/dsniff/}, 2000.
\newblock Accessed: 2023-11-24.

\end{thebibliography}

\bibliographystyle{unsrt}

\end{document}